\documentclass[12pt]{article}

\usepackage{amsfonts}
\usepackage{amsmath}
\usepackage{amssymb}
\usepackage{latexsym}

\usepackage[dvips]{graphicx}

\numberwithin{equation}{section}

\newcommand{\ba}{\begin{array}}
\newcommand{\ea}{\end{array}}
\newcommand{\be}{\begin{equation}}
\newcommand{\ee}{\end{equation}}
\newcommand{\bea}{\begin{eqnarray}}
\newcommand{\eea}{\end{eqnarray}}

\newcommand{\pg}{Painlev\'e-Gull\-strand} 

\newcommand{\massfunction}{{\mathcal{M}}}
\newcommand{\prmassfunction}{\massfunction}
\newcommand{\prmassoperator}{\widehat{\prmassfunction}}

\begin{document}
\begin{titlepage}
\renewcommand{\thefootnote}{\fnsymbol{footnote}}
\renewcommand{\baselinestretch}{1.3}
\medskip

\begin{center}
{\large {\bf Transgressing the horizons: Time operator in 
two-dimensional dilaton gravity}}
\\ \medskip {}
\medskip

\renewcommand{\baselinestretch}{1}
{\bf
Gabor Kunstatter${}^\dagger$
and 
Jorma Louko${}^\sharp$
\\}
\vspace*{0.50cm}
{\sl
${}^\dagger$ 
Department of Physics and Winnipeg Institute of
Theoretical Physics,\\ 
University of Winnipeg\\
Winnipeg, Manitoba, Canada R3B 2E9\\
{[e-mail: g.kunstatter@uwinnipeg.ca]}\\[5pt]
}
{\sl
${}^\sharp$ 
School of Mathematical Sciences,
University of Nottingham\\
Nottingham NG7 2RD, United Kingdom\\
{[e-mail: jorma.louko@nottingham.ac.uk]}\\ [10pt]
}
\vspace{2ex}
gr-qc/0608080\\[4ex]
Published in Phys.\ Rev.\ D {\bf 75}, 024036 (2007)\\[4ex]

{\bf Abstract}
\end{center}
We present a Dirac quantization of generic single-horizon black holes
in two-dimensional dilaton gravity. The classical theory is first
partially reduced by a spatial gauge choice under which the spatial
surfaces extend from a black or white hole singularity to a spacelike
infinity. The theory is then quantized in a metric representation,
solving the quantum Hamiltonian constraint in terms of (generalized)
eigenstates of the ADM mass operator and specifying the physical inner
product by self-adjointness of a time operator that is affinely
conjugate to the ADM mass. Regularity of the time operator across the
horizon requires the operator to contain a quantum correction that
distinguishes the future and past horizons and gives rise to a quantum
correction in the hole's surface gravity. We expect a similar quantum
correction to be present in systems whose dynamics admits black hole
formation by gravitational collapse.  {\small}
\vfill
\hfill  
Published version, January 2007 \\
\end{titlepage}

\section{Introduction}

Pure Einstein gravity in two spacetime dimensions is trivial, in the
sense that Einstein's vacuum field equations are satisfied by any
metric. Dynamically interesting two-dimensional gravity theories can
however be constructed by including suitable matter, and some such
two-dimensional theories are equivalent to a reduction of
higher-dimensional Einstein gravity to spherical
symmetry~\cite{vienna}. Quantization of two-dimensional gravity
theories thus presents an interesting problem, both as a dynamically
simplified setting for developing techniques that might be
generalizable to higher dimensions, as well as a quantization of the
spherically symmetric degrees of freedom of higher-dimensional
Einstein black holes. In particular, the macroscopic geometric
quantities that are associated with quantum black holes in the
semiclassical limit, such as the surface gravity of the horizon, are
all present in the two-dimensional setting. The quantization may
therefore be of interest from the semiclassical point of view even if
the fundamental building blocks of higher-dimensional gravity turn out
to be strings, spin networks or other pre-geometric
quantities~\cite{rovelli-review}.

In this paper we quantize a class of two-dimensional dilaton gravity
theories specified by a dilaton potential, under mild assumptions that
guarantee the classical solutions with positive ADM mass to be black
holes with a single, non-degenerate Killing horizon and suitable
asymptotics. This class of theories includes in particular
symmetry-reduced Einstein gravity in four or more spacetime
dimensions.

We first partially reduce the theory classically by a spatial gauge
choice \cite{geg1,dgk} that allows the spatial surfaces to extend from
a singularity to an infinity, crossing exactly one branch of the
horizon, and we choose boundary conditions that imply positivity of
the classical ADM mass, specify whether the singularity and horizon
are those of a black hole or a white hole, and prescribe the Killing
time evolution rate of the asymptotic ends of the spatial surfaces. We
then Dirac quantize this partially reduced theory in a metric
representation. The quantum Hamiltonian constraint is solved in terms
of eigenstates of the quantum ADM mass operator, and a class of
momentum-type quantum observables is constructed from classical
observables that are related to the time difference between the
asymptotic ends of the spatial surfaces. Transforming to a
representation that allows the ADM mass eigenstates to be treated as
non-normalizable states, we finally specify the inner product by
requiring that a particular momentum observable, affinely conjugate to
the ADM mass operator, is self-adjoint. The resulting spectrum of the
ADM mass operator is continuous and consists of the positive real
line.

The novel features of our quantum theory reside in the momentum
observables. The classical momentum observables are constructed to be
regular across the horizon that the spatial surfaces cross. As a
consequence, when evaluated across the other horizon, they pick up an
imaginary contribution inversely proportional to the hole's surface
gravity.  The corresponding quantum momentum observables are similarly
constructed to be regular across the horizon that the spatial surfaces
cross. When evaluated across the other horizon, they also turn out to
pick up an imaginary contribution, and this contribution differs from
that of the corresponding classical observable by a factor that
approaches unity for masses much larger than Planck mass but is
significantly smaller than unity near Planck mass and vanishes below
Planck mass. The singular contributions in the momentum observables
thus provide a definition of the 
inverse surface gravity operator in the quantum theory, with
significant quantum corrections at the Planck scale. The presence of
such quantum corrections can be understood as a consequence of the
fluctuations that our Dirac quantization of the Hamiltonian constraint
allows around the classical Hamiltonian constraint surface.

While the dynamical content of the system is limited in that the
classical theory has no local propagating degrees of freedom
\cite{vienna,exact,observables,birkhoff,%
kas-thie1,kas-thie2,kuchar,varadarajan,vara-kruskal}, 
we expect a number of the
features of the quantum theory to be generalizable upon inclusion of
matter that gives the system local
dynamics~\cite{dgk,krv,hw1,hw2,hw3}. In particular, we expect the
definition of regular quantum observables across the horizon to
survive. Also, as our foliation extends to the singularity of the
eternal hole, it may be possible in the presence of matter to
introduce boundary conditions that allow the study of singularity
formation in the quantum theory~\cite{dgk}.

The paper is organized as follows. The partially reduced classical
theory is presented in Section~\ref{sec:classical-rev}, and the theory
is quantized in Section~\ref{sec:quantization}. The inverse surface
gravity operator is constructed in
Section~\ref{sec:cross-qhor}. Section \ref{sec:conclusions} contains a
summary and a discussion.

\section{Classical theory}
\label{sec:classical-rev}

\subsection{Action and solutions}
\label{subsec:action-and-solutions}

We work with the action
\be
S[g,\phi]=\frac{1}{2G}\int d^2x \, \sqrt{-g}
\, \left(\phi R(g)+ \frac{V(\phi)}{l^2}\right) , 
\label{dgaction}
\ee
which is, up to conformal reparametrizations of the metric, the most
general two-dimensional second order, diffeomorphism invariant action
involving a metric $g_{\mu\nu}$ and a scalar~$\phi$ 
\cite{vienna,exact,observables}.
$l$~is a positive constant of dimension length and $G$ is the
two-dimensional Newton's constant. We do not need to fix the
physical dimension of~$G$, but since $GS$ is dimensionless, the
physical dimensions of $G$ and Planck's constant~$\hbar$, to be
introduced in Section~\ref{sec:quantization}, are related so that
$\hbar G$ is dimensionless.

The
action (\ref{dgaction}) can be obtained 
from a class of 
gravitational actions in $2+n$ dimensions by reduction to
the spherically symmetric ansatz 
\be
ds^2_{2+n} = \frac{ds^2}{j(\phi)} + r^2 d\Omega^2_{n} , 
\label{eq:spherical-ansatz}
\ee
where $n\ge2$, $d\Omega^2_{n}$ is the line element on unit~$S^{n}$,
$ds^2$ is the two-dimensional line element that appears
in~(\ref{dgaction}), $j(\phi)$ satisfies 
$dj/d\phi = V(\phi)$ and the area-radius $r$ is related to $\phi$ by
$\phi= {(r/l)}^n$ and $dj/d\phi = V(\phi)$.  The $(2+n)$-dimensional
action depends on the choice of the potential $V$ and equals
Einstein's action in the special case $V=\phi^{-1/n}$
\cite{vienna,exact,observables}. 

As one may expect from the special case of symmetry-reduced Einstein
gravity, the action (\ref{dgaction}) obeys a Birkhoff
theorem~\cite{birkhoff}. Assuming that $V(\phi)$ is nowhere vanishing,
the theorem states that the vector 
\be
k^\mu= \frac{1}{\sqrt{-g}}\epsilon^{\mu\nu}\partial_\nu \phi 
\label{eq:killing-vector}
\ee
is nonvanishing and Killing on every classical solution. Using $\phi$
as one of the coordinates, the solution can then be written in the
Schwarzschild-like form
\be
ds^2
= 
- {\bigl[ j(\phi) -2lGM \bigr]} dt_s^2
+{\bigl[ j(\phi) -2lGM \bigr]}^{-1}  l^2 d\phi^2 , 
\label{eq:schwarzschild-metric}
\ee
where $t_s$ is the Schwarzschild time coordinate, the Killing vector
(\ref{eq:killing-vector}) equals $\partial_{t_s}$ and the integration
constant $M$ is the ADM mass. Note that the combination $lGM$ is
dimensionless. From now on we assume $M>0$.

\begin{figure}[t]
\centering
\includegraphics[width=0.8\textwidth]{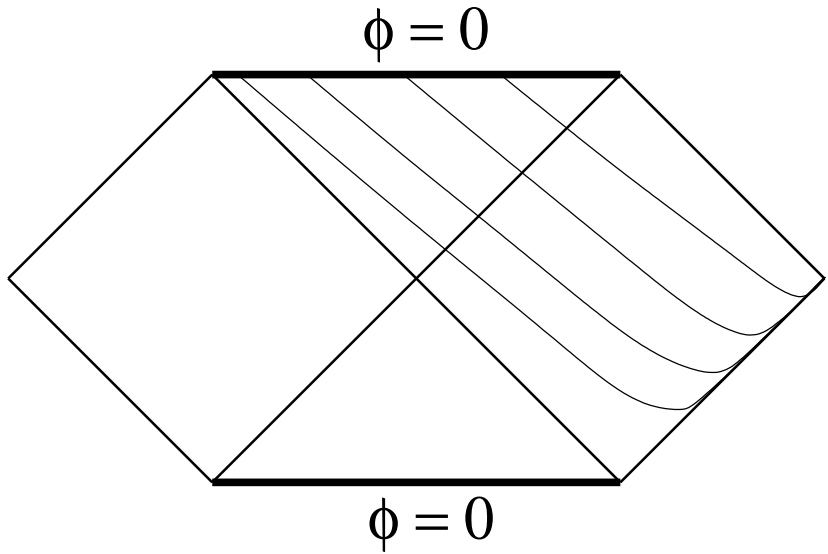}
\caption{Conformal diagram of the extended spacetime 
(\ref{eq:schwarzschild-metric}) with $M>0$, in the case of infinite 
$\int^\infty {\bigl[j(\phi)\bigr]}^{-1} d\phi$ (which implies that the
null infinities are distinct from the spacelike infinities). 
The thin lines show surfaces of constant
\pg{} $T$ (\ref{P-G-metric}) with $\epsilon=1$, 
assuming finite 
$\int_0 {\bigl[j(\phi)\bigr]}^{-1/2} d\phi$ (which determines the
asymptotics near the singularity) and infinite 
$\int^\infty {\bigl[j(\phi)\bigr]}^{-3/2} d\phi$ (which
determines the asymptotics near infinity).
The diagram for $\epsilon =-1$ is obtained by up-down inversion.}
\label{fig:conformal}
\end{figure}

We assume the potential $V(\phi)$ to be positive
and its
small $\phi$ behaviour to be such that $j(\phi)$ may be defined as
\be
j(\phi) := \int_0^\phi d{\tilde\phi}
\,V({\tilde\phi}) , 
\label{eq:j-def}
\ee 
with the consequence that $j(\phi)\to0$ as $\phi\to0$. These
assumptions hold in particular for symmetry-reduced Einstein
gravity. It follows that the $(2+n)$-dimensional metric
(\ref{eq:spherical-ansatz}) is generically singular at $\phi=0$, and
the two-dimensional metric (\ref{eq:schwarzschild-metric}) is
generically singular at $\phi=0$ for a range of theories, including
symmetry-reduced Einstein gravity. We therefore regard $\phi=0$ as a
singularity that is not part of the spacetime. At $\phi\to\infty$, we
assume that $j(\phi)$ grows without bound but so slowly that
$\int^\infty {\bigl[j(\phi)\bigr]}^{-1/2} d\phi$ is infinite. Again,
this holds for symmetry-reduced Einstein gravity. It follows that the
metric (\ref{eq:schwarzschild-metric}) has at $\phi\to\infty$ an
infinity, whose causal properties in terms of the null and spacelike
infinities depend on whether $\int^\infty {\bigl[j(\phi)\bigr]}^{-1}
d\phi$ is finite or infinite. The global structure of the spacetime
can be found by standard techniques~\cite{haw-ell}. There is precisely
one Killing horizon, which is bifurcate and located at
$j(\phi)=2lGM$~\cite{wald-smallbook}. The Killing vector
$\partial_{t_s}$ is timelike in the exterior regions, where $j(\phi) >
2lGM$, and spacelike in the black and white hole regions, where $0 <
j(\phi) < 2lGM$. Figure \ref{fig:conformal} shows a conformal diagram
of the case in which $\int^\infty {\bigl[j(\phi)\bigr]}^{-1} d\phi$ is
infinite.

We are interested in foliations that extend from $\phi=0$ to an
infinity at $\phi\to\infty$ and are regular across the
horizon. A~convenient example are the
\pg{} (PG) 
coordinates $(T,Y)$ \cite{painleve,gullstrand},
related to the
Schwarzschild coordinates (\ref{eq:schwarzschild-metric}) 
by 
\begin{subequations}
\bea
dY &=& \frac{l d\phi}{j(\phi)} , 
\\
dT &=& dt_s + \epsilon \, 
{\sqrt{\frac{2lGM}{j(\phi)}}}{\frac{l d\phi}{j(\phi)- 2lGM}} , 
\eea
\end{subequations}
where $\epsilon = \pm1$. 
The metric reads 
\be
ds^2= j(\phi)\left[-dT^2 
+ \left(dY + \epsilon  \, 
\sqrt{\frac{2lGM}{j(\phi)}} \, dT \right)^{\!\!2} 
\right] 
\label{P-G-metric}
\ee
and is clearly regular across the horizon. $\epsilon=1$ (respectively
$\epsilon = -1$) gives the ingoing (outgoing) PG metric, which
covers the black (white) hole region and one exterior region. The
asymptotic behaviour of the constant $T$ surfaces at $\phi\to0$ and
$\phi\to\infty$ depends on the asymptotic behaviour
of~$j(\phi)$. Figure \ref{fig:conformal} shows a sketch of these 
surfaces in the
case of finite 
$\int_0 {\bigl[j(\phi)\bigr]}^{-1/2} d\phi$ but infinite 
$\int^\infty {\bigl[j(\phi)\bigr]}^{-3/2} d\phi$, which occurs
in symmetry-reduced Einstein gravity in four and
five spacetime dimensions.

\subsection{Hamiltonian analysis}
\label{subsec:Hamiltonian} 

For the Hamiltonian analysis, we parametrize the metric as 
\be
ds^2=e^{2\rho}\left[-\sigma^2 dt^2+{(dx+N dt)}^2 \, \right] , 
\label{adm-decomposition}
\ee
where the rescaled lapse $\sigma$ and rescaled shift $N$ will play the
role of Lagrange multipliers.  
From the action (\ref{dgaction}) we find that the 
momenta conjugate to $\phi$ and $\rho$ are
\begin{subequations}%
\label{coord.eom}%
\bea
\Pi_\phi &=& \frac{1}{G \sigma} (N\rho'+N' - \dot{\rho}) , 
\\
\Pi_\rho &=& \frac{1}{G \sigma}(N\phi'-\dot{\phi}) , 
\label{eq:pi-rho-def}
\eea
\end{subequations}
where dot denotes derivative with respect to $t$ and prime denotes
derivative with respect to~$x$. 
The Hamiltonian action can be found by standard techniques 
\cite{Dir,Hen} 
and reads 
\be
S = \int dt \, dx \left( \Pi_\rho\dot{\rho} + \Pi_\phi\dot{\phi} \right) 
- \int dt \, H , 
\label{ham-action}
\ee
where the total Hamiltonian is
\be
H=\int dx \left(\sigma\mathcal{G} +N\mathcal{F}\right)
+ H_B , 
\label{hamiltonian}
\ee
the Hamiltonian constraint $\mathcal{G}$ 
and the momentum constraint $\mathcal{F}$ 
are given by%
\begin{subequations}%
\label{eq:ham-and-diffeo-constraints}%
\bea
G \mathcal{G}&:=&-G^2 \Pi_\rho\Pi_\phi+\phi''
-\phi'\rho'-\frac{1}{2 l^2} e^{2\rho} \, V(\phi),
\label{ham-constraint}
\\
\mathcal{F}&:=&\rho'\Pi_\rho-\Pi_\rho'+\phi'\Pi_\phi , 
\label{diffeo-constraint}
\eea
\end{subequations}
and $H_B$ consists of boundary terms evaluated at the (asymptotic)
upper and lower ends of the range of $x$.

The Hamiltonian equations of motion are the
constraint equations $\mathcal{G} = 0 = \mathcal{F}$ enforced by the
Lagrange multipliers, the momentum evolution equations%
\begin{subequations}%
\label{momenta-eom}%
\bea
G\dot{\Pi}_\phi&=& - \sigma'' - (\sigma\rho')' +\frac{\sigma}{2 l^2} 
e^{2\rho} \, \frac{dV}{d\phi} +(NG\Pi_\phi)' , 
\\
G\dot{\Pi}_\rho&=&(GN\Pi_\rho)' -(\sigma\phi')' 
+\sigma e^{2\rho} \, \frac{V(\phi)}{l^2} , 
\eea
\end{subequations} 
and the relations~(\ref{coord.eom}). To obtain these 
equations of motion from the action~(\ref{ham-action}), 
one needs to specify the boundary conditions and $H_B$ so
that the boundary terms in the variation of the action vanish. We
shall address this issue within the partially reduced theory in
Subsection~\ref{subsec:partially-reduced}.

\subsection{Spacetime reconstruction with the \pg{} time}
\label{subsec:reconstruction}

In this subsection we reconstruct from the canonical data $(\rho,
\phi, \Pi_\rho, \Pi_\phi)$ the spacetime and the location of the 
spacelike surface on which the canonical data is defined. We follow
closely Kucha\v{r}'s analysis of spherically symmetric Einstein
gravity in four dimensions~\cite{kuchar}, but we specify the
location of the surface in terms of the PG time
$T$~(\ref{P-G-metric}), rather than in terms of the Schwarzschild time
$t_s$~(\ref{eq:schwarzschild-metric}).
This will enable us to discuss the
regularity of the horizon-crossing in the quantum theory in
Section~\ref{sec:quantization}.

To begin, we define the mass function $\massfunction$ by 
\be
\massfunction 
:= 
\frac{1}{2lG}
\left\{ e^{-2\rho} \! \bigl[{l^2 G^2 \Pi^2_\rho}- l^2(\phi')^2 \bigr]+
j(\phi) \right\} . 
\label{mass-observable}
\ee
Differentiating with respect to $x$ 
and using~(\ref{eq:ham-and-diffeo-constraints}), we
find 
\be
\massfunction' = 
- l e^{-2\rho} \left(\phi'\mathcal{G} + G\Pi_\rho \mathcal{F}\right) . 
\label{eq:Mprime}
\ee
When the constraints hold, $\massfunction$ is therefore independent
of~$x$, and when all the equations of motion hold, $\massfunction$ is
also independent of~$t$. Comparison with
(\ref{eq:schwarzschild-metric}) or (\ref{P-G-metric}) shows that on a
classical solution $\massfunction$ is equal to the ADM mass~$M$.

To find the location of the surface in the spacetime, we look for
a coordinate transformation%
\begin{subequations}%
\label{eq:Tphi-transf}%
\bea
T&=&T(x,t)  , 
\\
\phi&=&\phi(x,t) , 
\eea 
\end{subequations} 
that brings the metric (\ref{adm-decomposition}) 
to the form 
\be
ds^2=j(\phi)\left[-dT^2 + (dY + F dT)^2\right] , 
\label{p-g-metric}
\ee
where
\be
dY = \frac{l d\phi}{j(\phi)}  
\label{eq:Y-def}
\ee 
and $F$ is initially unspecified. When the field equations hold, $F$
will turn out to be related to the ADM mass as shown
in~(\ref{P-G-metric}).

Differentiating (\ref{eq:Tphi-transf}) yields%
\begin{subequations}%
\label{dTdphi}%
\bea
dT&=&\dot{T}dt + T' dx , 
\\
d\phi&=&\dot{\phi}dt+\phi'dx . 
\eea
\end{subequations}
Substituting (\ref{dTdphi}) in 
(\ref{p-g-metric})
and comparing with~(\ref{adm-decomposition}), we obtain%
\begin{subequations}%
\label{eq:rec1}%
\bea
e^{2\rho}
&=& 
j(\phi)
\left[
A^2 -{(T')}^2
\right] , 
\label{conformal-factor}
\\
e^{2\rho} \bigl( \sigma^2-N^2 \bigr)
&=& j(\phi) \left( \dot{T}^2
- B^2
\right) , 
\label{eq:mixed}
\\
e^{2\rho}N
&=& 
j(\phi)
\left(
A B  -T'\dot{T}
\right) , 
\label{shift}
\eea
\end{subequations}
where%
\begin{subequations}%
\label{def-AB}%
\bea
A &:=& \frac{l \phi'}{j}+FT' , 
\label{def-A}
\\
B &:=& \frac{l\dot{\phi}}{j}+F\dot{T} . 
\eea
\end{subequations}
Solving (\ref{eq:rec1}) for $N$ and~$\sigma$, we find%
\begin{subequations}%
\label{eq:N-and-sigma-solved}%
\bea
N 
&=& \frac{A B  -T'\dot{T}}{A^2-{(T')}^2} ,
\\
\sigma 
&=& \frac{A\dot{T}- B T'}{A^2-{(T')}^2} . 
\label{eq:sigma-solved}%
\eea
\end{subequations}
Note that the denominators in (\ref{eq:N-and-sigma-solved}) are
positive because of~(\ref{conformal-factor}). To arrive
at~(\ref{eq:sigma-solved}) from (\ref{eq:mixed}), we have chosen the
sign of the square root so that $\sigma$ has the same sign as ${\dot
T}$ when $T'=0$. Assuming the metric to be invertible and both $T$ and
$t$ to increase towards the future, it then follows by continuity that
$\sigma$ is everywhere positive.

So far no field equations have been used. To proceed, we
substitute (\ref{eq:N-and-sigma-solved})
in~(\ref{eq:pi-rho-def}). Writing 
$\phi'$ and
${\dot \phi}$ 
in terms of $A$ and $B$ from~(\ref{def-AB}), 
we find that a cancellation occurs and allows the 
result to be written as 
\be
l G \Pi_\rho 
= 
j(\phi)(A F - T') . 
\label{eq:pirho-inter}
\ee
Eliminating $A$ from 
(\ref{def-A}) and (\ref{eq:pirho-inter}) yields
\be
T' = \frac{l (F\phi' - G\Pi_\rho)}{j(1-F^2)} . 
\label{eq:Tprime}
\ee
To find~$F$, we substitute (\ref{eq:Tprime}) in
(\ref{conformal-factor}) and (\ref{def-A}) and eliminate~$A$. Using 
(\ref{mass-observable}), we find 
\be
jF^2  = 2 l G \massfunction , 
\label{F-solution}
\ee
whose two solutions are 
\be
F= \pm \sqrt{\frac{2 l G \massfunction}{j}} . 
\label{eq:Fsol1} 
\ee
Collecting, we finally obtain 
\be
T' = \frac{l}{j- 2lG \massfunction}
\left(  
- G\Pi_\rho
\pm \sqrt{ \frac{2lG \massfunction}{j}}
\, \phi'
\right) . 
\label{eq:Tprimesol1}
\ee

To summarise, equations (\ref{mass-observable}), (\ref{eq:Fsol1}) and
(\ref{eq:Tprimesol1}) specify both the spacetime and the location of
the surface in the spacetime. When the field equations hold,
$\massfunction$ (\ref{mass-observable}) is the ADM mass, and
comparison of (\ref{P-G-metric}) with (\ref{p-g-metric}) and
(\ref{eq:Fsol1}) shows that for the upper (respectively lower) sign,
$T$ in (\ref{eq:Tprimesol1}) is the ingoing (outgoing) PG time. The
embedding of the surface in the spacetime is determined by the
canonical data by integrating (\ref{eq:Y-def})
and~(\ref{eq:Tprimesol1}), up to the isometries generated by the
Killing vector $\partial/\partial T$. Note that the first term in
(\ref{eq:Tprimesol1}) arises from the Schwarzschild time $t_s$
(\ref{eq:schwarzschild-metric})
\cite{observables,kuchar}
and the second term arises from the transformation to the 
PG time. Note also from (\ref{mass-observable})
that the zero in the denominator in (\ref{eq:Tprimesol1}) at the
horizon is cancelled by a zero in the numerator to give a finite limit
when the sign of $\Pi_\rho$ is such that the surface crosses
the horizon that the PG coordinates cover.

Although the spacetime interpretation of $T'$ (\ref{eq:Tprimesol1})
relies on the field equations, equation (\ref{eq:Tprimesol1}) can be
understood to define $T'$ as a function on the phase space
independently of the field equations
\cite{observables,kuchar}. 
We shall return to this in Subsection
\ref{subsec:partially-reduced} after having performed a partial
reduction and specified the boundary conditions.

\subsection{Partial reduction}
\label{subsec:partially-reduced}

The Hamiltonian action (\ref{ham-action}) contains two constraints,
the Hamiltonian constraint $\mathcal{G}$ and the spatial diffeomorphism
constraint~$\mathcal{F}$. We now
eliminate $\mathcal{F}$ by a spatial gauge condition that
fixes $\phi'$ to a given function of~$\phi$. For concreteness, we
focus on the gauge \cite{geg1}
\be
l\phi' - j(\phi) =0 , 
\label{eq:gauge-condition}
\ee
and postpone the discussion of other choices to 
Section~\ref{sec:conclusions}.

As the Poisson bracket of $\mathcal{F}$ and the 
left-hand side of (\ref{eq:gauge-condition}) is nonzero, the
gauge condition (\ref{eq:gauge-condition}) 
is admissible~\cite{geg1}. Substituting
(\ref{eq:gauge-condition}) in the action~(\ref{ham-action}), using
(\ref{eq:Mprime}) and introducing the rescaled lapse
${\tilde\sigma}$ by
\be
\tilde{\sigma} := \frac{\sigma e^{2\rho}}{j} , 
\ee
we obtain the action 
\be
S = \int dt \, dx \, 
\bigl( 
\Pi_\rho\dot{\rho}
+ {\tilde{\sigma}} \prmassfunction'  
\bigr)
+ S_B , 
\label{ham-red-action}
\ee
where $S_B$ is a boundary action, to be specified shortly, and the
mass function $\prmassfunction$ is now given by 
\be
\prmassfunction := 
\frac{1}{2lG}
\bigl[ e^{-2\rho} \! \bigl({l^2G^2 \Pi^2_\rho}
- j^2 \bigr)+ j 
\bigr] . 
\label{eq:prmassfunction-def}
\ee
For notational convenience, we suppress the 
$\phi$-dependence of~$j$ and continue to use for the mass function 
(\ref{eq:prmassfunction-def}) 
the same symbol as in the unreduced theory. 

The field equations read 
\be
\prmassfunction ' =  0  
\label{eq:pr-constraint-eq}
\ee
and%
\begin{subequations}%
\label{eq:tilde-dynamical}%
\bea
\dot{\rho}
&=& 
\tilde{\sigma}' e^{-2\rho} lG \Pi_\rho , 
\\
lG \dot{\Pi}_\rho
&=&
\tilde{\sigma}' e^{-2\rho}
\bigl({l^2G^2 \Pi^2_\rho}
- j^2 \bigr) . 
\label{eq:tilde-dynamical-Pi-rho}
\eea
\end{subequations}
If desired, $\Pi_\phi$ and $N$ can be recovered from the original
equations of motion (\ref{coord.eom}) and (\ref{momenta-eom}). In
particular, preservation of the gauge condition
(\ref{eq:gauge-condition}) yields
\be
N = 
\frac{\sigma lG \Pi_\rho}{j} = 
{\tilde\sigma}  e^{-2\rho} lG \Pi_\rho  . 
\label{dot-chi}
\ee

We are now in a position to address the boundary conditions at
$\phi\to\infty$ and $\phi\to0$. We choose for concreteness a falloff
that makes the foliation asymptotic to that of the PG coordinates
(\ref{p-g-metric}) at each end and postpone the discussion of other
choices to Section~\ref{sec:conclusions}. We also assume for
concreteness the large $\phi$ behaviour of $V(\phi)$ to be such that
there exists a positive constant $\beta$ for which the integral
\be
I_\beta^{+}(\phi) := 
\int_\phi^{\infty} 
d{\tilde\phi} {\bigl[j(\tilde\phi)\bigr]}^{-\beta - 3/2}
\ee
is finite. 
This holds for any potential that satisfies 
$V(\phi) > C \phi^{\gamma-1}$ at large~$\phi$, where $C$
and $\gamma$ are positive constants,
and holds therefore in particular for symmetry-reduced Einstein
gravity. 
To control the surfaces at small~$\phi$, we choose 
a positive constant $\alpha$ for which the integral 
\be
I_\alpha^{-}(\phi) := 
\int_0^{\phi} 
d{\tilde\phi} {\bigl[j(\tilde\phi)\bigr]}^{\alpha - 1/2}
\ee
is finite. The finiteness of (\ref{eq:j-def}) shows that 
a choice with 
$\alpha \ge1/2$ will work for all potentials. 

Given the positive constants $\alpha$ and~$\beta$, we impose at 
$\phi\to0$ the falloff
\bea
e^{2\rho}&=& j\bigl[1+ O( j^{\alpha})\bigr] , 
\nonumber\\
lG \Pi_\rho 
&=& 
\epsilon \sqrt{2lG M_0 j} 
\bigl[1+ O( j^{\alpha})\bigr] , 
\nonumber\\
\tilde{\sigma}&=& \sigma_0 
+ O 
\bigl( 
I_\alpha^{-} (\phi)
\bigr) , 
\label{eq:smallphi-falloff}
\eea
and at $\phi\to\infty$ the falloff
\bea
e^{2\rho}&=& j\bigl[1+ O( j^{-\beta - 1})\bigr] , 
\nonumber\\
lG \Pi_\rho 
&=& 
\epsilon \sqrt{2lG M_\infty j} 
\left[1+ O( j^{-\beta})\right] , 
\nonumber\\
\tilde{\sigma}&=&  \sigma_\infty 
+ O 
\bigl( 
I_\beta^{+} (\phi)
\bigr) , 
\label{eq:largephi-falloff}
\eea
where $\epsilon$ equals either $1$ or $-1$ and takes the same value in
both (\ref{eq:smallphi-falloff})
and~(\ref{eq:largephi-falloff}). $\sigma_0$, $\sigma_\infty$, $M_0$
and $M_\infty$ are independent of $x$ but may a priori depend
on~$t$. $M_0$~and $M_\infty$ are assumed positive. The $O$-terms may
depend on~$t$, and we assume that they can be treated under
algebraic manipulations and differentiation as series in 
powers of~$j$. 
It can be verified that this
falloff is consistent with the constraint (\ref{eq:pr-constraint-eq})
and preserved in time by the evolution
equations~(\ref{eq:tilde-dynamical}), where $\tilde\sigma$ remains
freely specifiable apart from the falloff. Note that the $O$-terms in
${\tilde\sigma}$ generate time evolution that affects the $O$-terms in
$\rho$ and $\Pi_\rho$ in precisely the order shown in
(\ref{eq:smallphi-falloff}) and~(\ref{eq:largephi-falloff}).  The
evolution equation (\ref{eq:tilde-dynamical-Pi-rho}) thus implies that
$M_0$ and $M_\infty$ are independent of~$t$, the constraint
(\ref{eq:pr-constraint-eq}) implies that $M_0$ and $M_\infty$ are
equal to each other, and it then follows from
(\ref{eq:prmassfunction-def}) that they are both equal to the ADM
mass. The foliation is at $\phi\to0$ and $\phi\to\infty$ asymptotic to
the
PG foliation~(\ref{P-G-metric}), with the values of $\epsilon$
matching. 
$\sigma_0$~and $\sigma_\infty$ remain
freely specifiable functions of~$t$, and they give the rate at which
the asymptotic PG times evolve with respect to~$t$. Finally, the action
(\ref{ham-red-action}) and its variation under these conditions can be
verified to be well-defined if we set
\be
S_B = - \int dt \, (\sigma_\infty M_\infty -\sigma_0 M_0) , 
\ee
where $\sigma_\infty$ and $\sigma_0$ are freely prescribable as
functions of $t$ but are considered fixed in the variation. Note that
the total action can be written in the alternative form
\be
S = \int dt \, dx 
\left( 
\Pi_\rho\dot{\rho}
- {\tilde{\sigma}}' \prmassfunction
\right) . 
\label{ham-red-action-alt}
\ee

Consider now observables (or
``perennials''~\cite{kuchar-winnipeg}). The mass function
$\prmassfunction$ (\ref{eq:prmassfunction-def}) has clearly a vanishing
Poisson bracket with the single remaining constraint and is hence an
observable. To find a second observable, we define
\be
\Pi_{\prmassfunction} := \frac{lG \Pi_\rho
-\epsilon \sqrt{2lG \prmassfunction j}}
{j- 2lG \prmassfunction} . 
\label{eq:Pi-prmassf-def}
\ee
The right-hand side of 
(\ref{eq:Pi-prmassf-def}) is not 
defined at the zeroes of the denominator, 
but if $\Pi_\rho$ has the same sign as in the
falloff region, it follows from (\ref{eq:prmassfunction-def}) 
that 
$\Pi_{\prmassfunction}$ 
can be written as 
\be
\Pi_{\prmassfunction} 
= 
\frac{\epsilon (j- e^{2\rho})}
{\sqrt{j^2 + e^{2\rho} ( 2lG\prmassfunction - j)}
+\sqrt{2lG\prmassfunction j}} , 
\label{eq:Pi-prmassf-alt}
\ee
which is nonsingular at the zeroes of the denominator
in~(\ref{eq:Pi-prmassf-def}). The phase space therefore contains a
neighbourhood of the classical solutions in which
$\Pi_{\prmassfunction}$ is well defined by~(\ref{eq:Pi-prmassf-def}),
supplemented by (\ref{eq:Pi-prmassf-alt}) at the zeroes of
the denominator. We restrict the attention to this neighbourhood. As
the notation suggests, 
$\Pi_{\prmassfunction}$ is conjugate to~$\prmassfunction$,
\be
\bigl\{
\prmassfunction(x) , \Pi_{\prmassfunction}(y)
\bigr\}
=\delta(x-y) . 
\label{eq:prmass-Pi-prmass-poisson}
\ee

From (\ref{eq:prmass-Pi-prmass-poisson}) it follows that
$\Pi_{\prmassfunction}$ in its own right is not an
observable. Consider, however the quantity
\be
P := \int dx \,  \Pi_{\prmassfunction}(x) , 
\label{eq:P-def}
\ee
where the convergence of the integral is guaranteed by the falloff 
(\ref{eq:smallphi-falloff})
and~(\ref{eq:largephi-falloff}). 
From (\ref{eq:prmass-Pi-prmass-poisson}) we find 
\be
\bigl\{
\prmassfunction(x) , P 
\bigr\}
= 1 . 
\label{eq:prmass-P-poisson}
\ee
If $\lambda(x)$ is the infinitesimal parameter of a gauge
transformation, vanishing at the upper and lower limits of~$x$, the
infinitesimal change in $P$ under this transformation reads 
\be
\left\{
P, \int dx \, \lambda'(x) \prmassfunction (x)
\right\} 
= - \int dx \, \lambda'(x) 
=0 . 
\ee
Hence $P$ is an observable. 


When the equations of motion hold, equations (\ref{eq:Tprimesol1}) and
(\ref{eq:gauge-condition}) show that $\Pi_{\prmassfunction} = -
T'$, where $T$ is the PG time, ingoing for $\epsilon=1$ and
outgoing for $\epsilon=-1$. In terms of the spacetime geometry,
$P$~is therefore equal to the difference of the PG times at the
left and right ends of the spatial surface. Note that this geometric
interpretation is consistent with the equation of motion for~$P$,
\be
\dot{P} 
= 
\left\{
P, \int dx \, {\tilde{\sigma}}'(x) \prmassfunction (x)
\right\} 
= \sigma_0 - \sigma_\infty . 
\ee

The fully reduced theory can be obtained by taking the spatially
constant value of $\prmassfunction$ as a new phase space
variable. Denoting this variable by $M$ and proceeding as
in~\cite{observables,kuchar}, we find the fully reduced action 
\be
S_{\mathrm{red}} = 
\int dt 
\left[
P \dot{M} - (\sigma_\infty -\sigma_0) M 
\right] . 
\label{eq:fully-reduced-action}
\ee
$P$ is therefore conjugate to the ADM mass in the fully reduced
theory.


\section{Quantization of the partially reduced theory}
\label{sec:quantization}

In this section we quantize the partially reduced theory of
Subsection~\ref{subsec:partially-reduced}. Following Ashtekar's
algebraic extension of Dirac
quantization~\cite{Ash1,Ash2}, 
we first find a vector space of solutions
to the quantum constraint and then determine the physical inner
product from the adjointness relations of a judiciously-chosen set of
quantum observables.

\subsection{Classical constraint}

We begin with some observations about the classical constraint. 

It is convenient to transform from the canonical pair $(\rho,
\Pi_\rho)$ to the pair $(X, P_X)$, where $X=e^\rho$ and
$P_X=e^{-\rho}\Pi_\rho$. The mass function
(\ref{eq:prmassfunction-def}) takes the form
\be
\prmassfunction = 
\frac{1}{2lG}
\left(
l^2 G^2 P_X^2
- \frac{j^2}{X^2} + j
\right) , 
\label{eq:prmassfunction-alt}
\ee
and the solutions to the classical constraint equation
(\ref{eq:pr-constraint-eq}) can be written as
\be
l^2 G^2 P_X^2 - \frac{j^2}{X^2} 
= 
2 l G M - j , 
\label{eq:class-X-constraint}
\ee
where the integration constant $M$ is the value
of~$\prmassfunction$, independent of the spatial
coordinate~$x$. The boundary conditions of Subsection
\ref{subsec:partially-reduced} imply that $M$ is positive.

For each~$x$, equation (\ref{eq:class-X-constraint}) can be understood
as the classical energy conservation equation of a particle moving on
the half-line of positive $X$ with the (true) Hamiltonian
\be
H := 
l^2 G^2 P_X^2 - \frac{j^2}{X^2} , 
\label{eq:true-Ham}
\ee
which consists of a conventional quadratic kinetic term and the
attractive potential well~$-j/\bigl(X^2\bigr)$. The value of the
energy is $2 lG M - j$, which is positive (respectively negative) for
those values of $x$ that in the spacetime are inside (outside) the
hole. 
We shall see that the oscillatory/exponential behaviour of the
solutions to the quantum constraint in Subsection
\ref{subsec:quantum-constraint} is in agreement with this classical
picture.

We note in passing that the Poisson bracket algebra of $H$ 
(\ref{eq:true-Ham}) and the functions
\be
D := \frac{X P_X}{2}  
\ \ , \ \ 
\\
K := \frac{X^2}{4 l^2 G^2} , 
\ee
at fixed $x$ is the $\mathfrak{o}(2,1)$ algebra,
\be
\{D,H\} = H ; \quad \{K,H\} = 2D; \quad \{K,D\} = K . 
\label{eq:HDK-brackets}
\ee
In particular, the first of the brackets in (\ref{eq:HDK-brackets}) is
equivalent to the observation that $H$ is scale invariant: Under the
scale transformation $(X, P_X) \to (\alpha X, P_X/\alpha)$, where
$\alpha$ is a positive constant, $H$ only changes by an overall
multiplicative factor. In terms of the spacetime geometry, $K =
e^{2\rho}$ is the conformal factor in the metric
(\ref{adm-decomposition}) and $D$ can be related to the expansion of
null geodesics~\cite{geg1}.  The potential interest in this
observation is that quantization of $H$, $D$ and $K$ forms the basis
of conformal quantum mechanics~\cite{alfaro-conformal}, and it has
been suggested that a near-horizon conformal symmetry could account
for black hole microstates and black hole
entropy~\cite{carlip-conf-horizon,solodukhin-conf-horizon}. 
There are however two obstacles to making
progress from this observation in the present context. First, the
classical $\mathrm{O}(2,1)$ symmetry generated by $H$, $D$ and $K$
cannot be promoted into a symmetry of conformal quantum mechanics ---
it develops an anomaly~\cite{quantum-conformal}. Second, as the
classical system still has one constraint, the phase space functions
functions $H$, $D$ and $K$ are not classical observables, and their
quantization by the methods of conformal quantum mechanics would
somehow need to accommodate a quantum version of the remaining
constraint. We shall not pursue this line further here.

\subsection{Quantum constraint}
\label{subsec:quantum-constraint}

We quantize in a representation in which the quantum states are
functionals of~$X(x)$. The operator substitution in this
representation at each $x$ is
\be
P_X \to -i 
\left( 
\frac{\hbar}{l} 
\right)
\frac{\partial}{\partial X} , 
\label{eq:op-subst}
\ee
where $\hbar$ is Planck's constant and the factor $1/l$ is required
for dimensional consistency because of the functional dependence
on~$x$. Suppressing~$x$, we promote the mass function
(\ref{eq:prmassfunction-alt}) into the mass operator
\be
\prmassoperator
:= 
\frac{1}{2lG}
\left(
- \hbar^2 G^2 
\frac{\partial^2}{\partial X^2}
- \frac{j^2}{X^2} +
j 
\right) . 
\label{eq:prmassoperator-def}
\ee
Note that the combination $\hbar G$ is dimensionless, 
as we observed in Subsection~\ref{subsec:action-and-solutions}. 
Following Dirac's procedure~\cite{Dir,Hen}, we then promote the 
classical constraint equation (\ref{eq:pr-constraint-eq}) 
into the quantum constraint equation 
\be
{
\bigl(
\prmassoperator
\bigr)
}'
\Psi =0 . 
\label{eq:pr-constraint-quantum}
\ee

We look for quantum states that are eigenstates of~$\prmassoperator$, 
\be 
\prmassoperator \Psi_M = M \Psi_M , 
\label{eq:eigenstate-equation}
\ee 
where the eigenvalue $M$ is independent of~$x$.
As the classical
boundary conditions of Subsection
\ref{subsec:partially-reduced} assume the ADM mass to be positive, 
we take $M>0$.
It is immediate from (\ref{eq:pr-constraint-quantum}) 
that $\Psi_M$ is annihilated by the quantum
constraint. Using~(\ref{eq:prmassoperator-def}),
equation (\ref{eq:eigenstate-equation}) reads 
\be
\left(
- \hbar^2  G^2 \frac{\partial^2}{\partial X^2} - \frac{j^2}{X^2}
\right) 
\Psi_M = ( 2 l G M-j ) \Psi_M . 
\label{schrodinger} 
\ee
Note that (\ref{schrodinger}) is the quantized version
of~(\ref{eq:class-X-constraint}). While (\ref{schrodinger}) is still a
functional differential equation in the variable~$X(x)$, the absence
of derivatives with respect to $x$ implies that the different spatial
points decouple, and we may separate the solution with the ansatz
\bea
\Psi_M \bigl(X(x)\bigr)
&
=& \prod_x 
\psi_M (X;x)
\nonumber\\ 
&:=& \exp 
\left\{ 
\int \frac{dx}{l} \ln \bigl[ \psi_M(X;x) 
\bigr] 
\right\} , 
\label{eq:PsiM-decomp}
\eea 
where the infinite product over $x$ is defined via the integral
expression. The factor $1/l$ in the integration measure is required
for dimensional consistency. $\psi_M(X;x)$ then satisfies
(\ref{schrodinger}) as an ordinary differential equation at each~$x$,
\be
\left(
-\hbar^2 G^2 \frac{\partial^2}{\partial X^2} - \frac{j^2}{X^2}
\right)
\psi_M(X;x) 
= 
( 2 l G M-j ) \psi_M(X;x) . 
\label{ode}
\ee

A solution to (\ref{ode}) for $2lGM -j \ne0$ is 
\be
\psi_M^{\nu}(X;x) := \omega^{-\nu} \sqrt{X} J_{\nu} (\omega X) , 
\label{solutions}
\ee
where $J_\nu$ 
is the Bessel function of the first kind \cite{Ab-Steg}
and
\bea
\omega^2 &=&  \frac{2 l G M-j}{\hbar^2 G^2} , 
\label{eq:omega-def}
\\
\nu^2 &=& \frac{1}{4} - \frac{j^2}{\hbar^2 G^2} . 
\label{eq:nusquared-def}
\eea
The branch point structure of $J_\nu$ implies that 
$\psi_M^{\nu}$ is 
independent of the sign taken in solving 
(\ref{eq:omega-def}) for~$\omega$. 
For $2 l G M-j=0$, we take $\psi_M^{\nu}$ to be given by the 
$\omega\to0$ limit of~(\ref{solutions}), 
$X^{\nu + 1/2}\big/\bigl[2^\nu \Gamma(\nu+1)\bigr]$, 
which again is a solution to~(\ref{ode}).
$\psi_M^{\nu}$~is then regular 
as a function of $x$ everywhere, including the zero of $2 l G M-j$. 

For $j \ne \hbar G/2$, the functions $\psi_M^{\nu}$ with the two
values of $\nu$ (\ref{eq:nusquared-def}) are linearly
independent. The case $j = \hbar G/2$ is special since $\nu=0$,
and if a linearly independent second solution to (\ref{ode}) were
desired, it could be given in terms of a Neumann
function~\cite{Ab-Steg}. For our purposes, $\psi_M^{\nu}$ will
suffice for all~$\nu$.

At $X\to\infty$, $\psi_M^{\nu}$ is oscillatory for $\omega^2>0$ and
exponentially increasing for $\omega^2<0$. If (\ref{ode}) were
interpreted as the time-independent Schr\"odinger equation for the
quantization of the classical Hamiltonian (\ref{eq:true-Ham}) in the
Hilbert space $L_2(\mathbb{R_+}, dX)$, the relevant solution for
$\omega^2 <0$ would therefore not be $\psi_M^{\nu}$ but instead the
exponentially decreasing linear combination proportional to $\sqrt{X}
K_{\nu} (\sqrt{-\omega^2} X)$, where $K_\nu$ is the modified Bessel
function of the second kind~\cite{Ab-Steg}. The possible negative
values of $\omega^2$ would be discrete and determined by the
self-adjointness boundary condition at $X\to0$~ (see Example 2.5.14 in
\cite{thirring}); in particular, for $\nu^2<0$ the spectrum of
$\omega^2$ would be unbounded from below with every choice of the
boundary condition. The relevant solution for $\omega^2>0$ would
similarly be determined by the self-adjointness boundary condition at
$X\to0$ and would coincide with $\psi_M^{\nu}$ only when $\nu^2\ge0$
and one of two special boundary conditions is chosen. In the
present context, however, there is no reason to relate the solutions
to $L_2(\mathbb{R_+}, dX)$, and we may continue to work with
$\psi_M^{\nu}$. A~quantum regularity condition that will be imposed in
Subsection \ref{subsec:q-observables} will in fact exclude linear
combinations of $\psi_M^{\nu}$ with the two signs of~$\nu$.

\subsection{Quantum observables}
\label{subsec:q-observables}

Recall that the classical observables $\prmassfunction$
(\ref{eq:prmassfunction-def}) and $P$ (\ref{eq:P-def}) induce a global
canonical chart on the fully reduced phase space. If $f$ is a
smooth function of a real variable, $f(\prmassfunction)$ and
$f(\prmassfunction) P$ are thus classical observables, and the set of
such observables is large enough to separate the fully reduced phase
space. In this subsection we define corresponding quantum observables
in the partially reduced quantum theory as linear operators on a
vector space annihilated by the quantum constraint. 

We begin with the `momentum' observables. As preparation, consider
$\Pi_{\prmassfunction}$~(\ref{eq:Pi-prmassf-def}). In terms of the
canonical pair $(X,P_X)$, we have
\be
\Pi_{\prmassfunction} 
= 
\frac{lG X P_X
-\epsilon \sqrt{2lG \prmassfunction j}}
{j- 2lG \prmassfunction} . 
\label{eq:Pi-prmassf-newchart}
\ee
We seek to define the corresponding operator
$\widehat{\Pi_{\prmassfunction}}$ on the mass eigenstates by 
\be
\widehat{\Pi_{\prmassfunction}} 
\, 
\psi_M^{\nu}
:=
\frac{ -i \hbar G 
\bigl( X \partial_X + \eta \bigr)
-\epsilon \sqrt{2lG M j}}
{j- 2lG M} 
\, 
\psi_M^{\nu} , 
\label{eq:Pi-prmass-redop}
\ee
where the factor ordering parameter $\eta$ may depend on $x$ but not
on~$M$. Since both 
$\Pi_{\prmassfunction}$ and $\psi_M^{\nu}$ are
regular as functions of $x$ across $2 l G M-j=0$, we postulate
also $\widehat{\Pi_{\prmassfunction}} \, \psi_M^{\nu}$ 
to be regular
as a function of $x$ across $2 l G M-j=0$. 
Using identity 9.1.27 of \cite{Ab-Steg} to write
(\ref{eq:Pi-prmass-redop}) as 
\be
\widehat{\Pi_{\prmassfunction}} 
\, 
\psi_M^{\nu}
= 
\frac{ -i \hbar G 
\bigl( \nu + \frac12 + \eta \bigr)
-\epsilon \sqrt{2lG M j}}
{j- 2lG M} 
\, 
\psi_M^{\nu} 
\, - \, 
i\frac{X \psi_M^{\nu+1}}{\hbar G} , 
\label{eq:Pi-prmass-redop-alt}
\ee
where the last term is always regular across $2lGM-j=0$, 
we see that this regularity condition implies 
\be
\eta 
= 
- \frac12
- \nu 
+ i \frac{\epsilon j}{\hbar G} . 
\label{eq:eta-raw}
\ee
We further postulate that $\eta$ remain bounded as $\hbar\to0$, as
expected of a factor ordering parameter. To achieve this, we choose
the sign of $\nu$ for $j > \hbar G / 2$ so that
\be
\nu = i \epsilon 
\sqrt{ \frac{j^2}{\hbar^2 G^2} - \frac{1}{4}} . 
\ee
We leave the sign of $\nu$ for $j < \hbar G / 2$ unspecified. 


Given the classical observable $f(\prmassfunction)P$, we now
define the corresponding operator $\widehat{fP}$ on the mass
eigenstates by
\be
\widehat{fP} := \int dx \,  
\widehat{\Pi_{\prmassfunction}} \, 
\widehat{f(\prmassfunction)}  \ . 
\label{eq:fP-on-Meigen}
\ee
A convenient phase choice for the mass eigenstates is
\bea
\Phi_M 
&:=& \prod_x 
E_M \psi_M^{\nu}
\nonumber\\ 
&:=& \exp 
\left[
\int \frac{dx}{l} \ln 
\bigl( E_M \psi_M^{\nu} \bigr) 
\right] , 
\label{eq:PhiM-def}
\eea
where 
\be
E_M := 
\exp\left\{
- i \frac{\epsilon}{\hbar G}
\left[
\sqrt{2lGMj} 
- j \ln
\left(
\sqrt{j} + \sqrt{2lGM} 
\right)
\right]
\right\} . 
\ee
We then find 
\bea
\widehat{fP} \, \Phi_M 
&=&
f(M) 
\int dx \, 
\widehat{\Pi_{\prmassfunction}} \Phi_M 
\nonumber
\\
&=&
f(M) \, \Phi_M 
\left(
\int dx \, 
\frac{ \widehat{\Pi_{\prmassfunction}} \, \Phi_M}
{\Phi_M }
\right)
\nonumber
\\
&=&
f(M) \, \Phi_M 
\left[
\int dx \, 
\frac{ \widehat{\Pi_{\prmassfunction}} \, \bigl( E_M \psi_M^{\nu} \bigr)}
{E_M \psi_M^{\nu} }
\right]
\nonumber
\\
&=&
f(M) \, \Phi_M 
\left[
\int dx \, 
\frac{ i (\hbar/l) \, \partial_M \bigl( E_M \psi_M^{\nu} \bigr)}
{E_M \psi_M^{\nu}}
\right]
\nonumber
\\
&=&
f(M) \, \Phi_M 
\left[
i \hbar \, \partial_M
\int \frac{dx}{l} 
\ln \bigl( E_M \psi_M^{\nu} \bigr)
\right]
\nonumber
\\
&=&
i \hbar \, f(M) \, \partial_M \Phi_M , 
\label{eq:fPhat-on-PhiM}
\eea
where we have used the identity 
\be
\widehat{\Pi_{\prmassfunction}} 
\, 
\bigl( E_M \psi_M^{\nu} \bigr) 
=  
i (\hbar/l) \, \partial_M
\bigl( E_M \psi_M^{\nu} \bigr) , 
\label{eq:PiMhat-x}
\ee
which follows by observing that $X\partial_X \bigl(
\omega^{\nu + (1/2)} \psi_M^{\nu} \bigr) =
\omega \partial_\omega \bigl( \omega^{\nu + (1/2)} \psi_M^{\nu} \bigr) 
= 
\tfrac12 M \partial_M \bigl( \omega^{\nu + (1/2)} \psi_M^{\nu}
\bigr)$. 

The `position' observables are straightforward: Given the classical
observable $f(\prmassfunction)$, we define the corresponding
quantum observable $\widehat{f}$ on the mass eigenstates by
\be
\widehat{f} \, \Phi_M := f(M) \Phi_M . 
\label{eq:fhat-on-PhiM}
\ee

To obtain an observable algebra that acts on a vector space, we extend
formulas (\ref{eq:fPhat-on-PhiM}) and (\ref{eq:fhat-on-PhiM}) to
define the action of the momentum and position observables on more
general functions of the variable $X(x)$ and the parameter~$M$. Given
this action, we then build the vector space $V :=
\mathcal{A} \bigl( \mathrm{span}\{ \Phi_M \}
\bigr)$, where 
$\mathcal{A}$ is the algebra generated by the momentum and position
observables. $V$~carries by construction a representation
of~$\mathcal{A}$, and viewing the derivative in
(\ref{eq:fPhat-on-PhiM}) as the limit of a differential quotient
provides by linearity a sense in which $V$ is annihilated by the
quantum constraint. 
One might thus attempt to 
define a quantum theory by introducing an inner product on~$V$, or
possibly on some subspace obtained by replacing $\mathrm{span}\{
\Phi_M
\}$ by a suitable subspace and
$\mathcal{A}$ by a suitable subalgebra. 
A~quantum theory of this kind would be expected to contain
mass eigenstates as normalizable states. 
While discrete black hole spectra have been encountered in a number of
approaches (see 
\cite{bekenstein1,stro-vafa,loop-entropy,louko-makela,bar-kunstatter}
for a small selection and \cite{louko-makela} for a more extensive
bibliography), we shall modify the representation in a way that will
lead to a continuous mass spectrum.

\subsection{Physical Hilbert space}

We look for a quantum theory in which the spectrum of
$\prmassoperator$ is continuous and consists of the positive
half-line. While the mass eigenstates $\Phi_M$ do then not exist as
normalizable states, one expects there to exist a spectral
decomposition in which any sufficiently well-behaved function
$\alpha:\mathbb{R_+}
\to
\mathbb{C}$ defines a normalizable state by the map 
\be
\alpha \mapsto \int_0^\infty \frac{dM}{M} \, \alpha(M) \, \Phi_M . 
\label{eq:spectral-decomp}
\ee
The factor $1/M$ in the integration measure is a convention that
will simplify what follows. If formula (\ref{eq:spectral-decomp})
holds in a sense that allows integration by parts without boundary
terms, the representation of $\mathcal{A}$ given by
(\ref{eq:fPhat-on-PhiM}) and (\ref{eq:fhat-on-PhiM}) then induces on
the space of the sufficiently well-behaved functions 
the representation%
\begin{subequations}%
\label{eq:indrep}%
\bea
\bigl(\widehat{f} \, \alpha \bigr)(M) 
&=& 
f(M)\alpha(M) , 
\\
\bigl(\widehat{fP} \, \alpha\bigr) (M) 
&=& 
- i \hbar \, M 
\frac{d}{dM} 
{\left[ \frac{f(M)}{M} \alpha(M) \right]} . 
\eea
\end{subequations}


To build a quantum theory with these properties, we adopt
(\ref{eq:indrep}) as the \emph{definition\/} of the
$\mathcal{A}$-action on the
space $\mathcal{C}_0^\infty(\mathbb{R_+})$ 
of smooth compactly-supported functions $\alpha: \mathbb{R_+}
\to \mathbb{C}$. 
This gives in particular the commutators%
\begin{subequations}%
\label{eq:can-commutators}%
\bea
\bigl[ \prmassoperator, \widehat{P} \, \bigr] 
&=& i\hbar ,
\label{eq:can-M-P-commutator} 
\\
\bigl[ \prmassoperator, \widehat{\prmassfunction P} \, \bigr] 
&=& i\hbar \, \prmassoperator . 
\label{eq:can-M-MP-commutator} 
\eea
\end{subequations}

We look on 
$\mathcal{C}_0^\infty(\mathbb{R_+})$
for an inner product $(\,\cdot\,,\cdot\,)$ 
of the form
\be
\bigl( \alpha_2, \alpha_1 \bigr) 
= 
\int_0^\infty dM \mu(M) \, \overline{\alpha_2(M)} \, \alpha_1(M) , 
\ee
where the overline denotes complex conjugation and the positive weight
function $\mu$ is to be specified. For any real-valued
function~$f$, the corresponding operator $\widehat{f}$ is then 
essentially self-adjoint. In particular, 
$\prmassoperator$ is essentially self-adjoint and has
spectrum~$\mathbb{R_+}$. 
From this and the commutator (\ref{eq:can-M-P-commutator}) 
it follows that $\widehat{P}$ 
does not have self-adjoint extensions for any
$\mu$~\cite{klau-aff,klau-asl}. 
However, the affine commutation
relation (\ref{eq:can-M-MP-commutator}) shows that
$\widehat{\prmassfunction P}$ can be made self-adjoint. 
Requiring 
$\widehat{\prmassfunction P}$ to be symmetric, $\bigl(
\alpha_2, \widehat{\prmassfunction P} \, \alpha_1
\bigr) 
= 
\bigl( 
\widehat{\prmassfunction P} \, \alpha_2, \alpha_1
\bigr)$, gives for $\mu$ a differential equation whose solution is 
$\mu(M) = c/M$, where the constant $c$ can be set to $1$ without loss
of generality. Completion of $\mathcal{C}_0^\infty(\mathbb{R_+})$ in
this inner product yields the Hilbert space 
$L_2 (\mathbb{R_+}, dM/M)$, 
on which $\widehat{\prmassfunction P}$ is essentially
self-adjoint 
\cite{thirring,klau-aff,klau-asl}. 
The mass eigenstates in the spectral decomposition
(\ref{eq:spectral-decomp}) can be understood as non-normalizable
states that satisfy
\be
\bigl( \Phi_M, \Phi_{M'} \bigr) = M \delta(M,M') , 
\ee
where $\delta$ is the Dirac delta-function. 

The algebra $\mathcal{A}$ is by construction represented on the
dense domain $\mathcal{C}_0^\infty(\mathbb{R_+}) 
\subset 
L_2 (\mathbb{R_+}, dM/M)$
and provides thus a large class of observables for the quantum
theory.

\section{Crossing the quantum horizon}
\label{sec:cross-qhor}

The observables of the classical Hamiltonian theory contain
information about the ADM mass of the spacetime and about the relative
location of the asymptotic ends of the spatial surface, but no
information about the spatial surface between its asymptotic
ends. Similarly, operators in the quantum observable algebra
$\mathcal{A}$ come with a geometric interpretation in terms of the ADM
mass and the relative location of the asymptotic ends of the spatial
surfaces, but not in terms of the local spacetime geometry. While this
is to be expected, owing to the absence of local propagating degrees
of freedom in the classical theory, we now show that the time-asymmetry
built into the theory provides a way to introduce a quantum operator
that is related to the surface gravity of the horizon.

Recall first that the spatial surfaces in the classical theory were
chosen to extend from a singularity to an infinity, crossing the black
hole horizon for $\epsilon=1$ and the white hole horizon for
$\epsilon=-1$. The classical momentum observables of the form
$f(\prmassfunction)P$ depend explicitly on $\epsilon$ as seen
from~(\ref{eq:Pi-prmassf-def}), and they have a geometric
interpretation in terms of the ADM mass and the PG time difference
between the left and right ends of the spatial surface. 

Consider the classical theory with given~$\epsilon$, and denote $P$ in
this theory by $P^\epsilon$ to explicitly indicate its dependence
on~$\epsilon$. Suppose that we attempt to introduce in this theory
momentum observables of the form~$fP^{-\epsilon}$. Proceeding for the
moment formally, we obtain
\be
f(\prmassfunction)P^{-\epsilon}
= 
f(\prmassfunction)P^{\epsilon}
+ 
f(\prmassfunction) \int
dx \, 
\frac{2\epsilon \sqrt{2lG \prmassfunction j}}
{j- 2lG \prmassfunction} . 
\label{eq:fPdiff}
\ee
The integral in (\ref{eq:fPdiff}) is clearly convergent at the lower
end of~$x$. The integral is convergent at $x\to\infty$ if $\int^\infty
{\bigl[j(\phi)\bigr]}^{-3/2} d\phi$ is finite, which means
geometrically that the surfaces of constant PG time asymptote to
surfaces of constant Schwarzschild time. We assume this to be the case
here and return to the question in Section~\ref{sec:conclusions}.

Let $\prmassfunction$ take the spatially constant value~$M$. The
integral in (\ref{eq:fPdiff}) is then singular across $j = 2lGM$, the
geometric reason being that the outgoing (respectively ingoing) PG
time tends to $\infty$ ($-\infty$) upon approaching the black (white)
hole horizon from the exterior. However, the integral is well defined
in the principal value sense~\cite{kuchar}, as well as in a contour
integral sense
\cite{observables} provided one specifies the half-plane in $x$ to
which the contour is deformed. 
If the contour circumvents the pole
in the upper (lower) half of the complex $x$ plane, and if $f$ is
real-valued, we thus obtain
\be
\mathrm{Im} \,  
\bigl[fP^{-\epsilon} \bigr] 
= 
\mp 
\frac{\epsilon \pi f(M)}{\kappa(M)} , 
\label{eq:fPminus-imagpart-solution}
\ee
where $\kappa(M)$ is the surface gravity of the horizon, given by
$\kappa = {(2l)}^{-1} V(\phi_H)$, with $\phi_H$ denoting the value of
$\phi$ at the horizon. In this sense, the inverse surface gravity of
the horizon can be recovered from a controllable singularity in the
observable~$fP^{-\epsilon}$.

We note in passing that replacing $P^{\epsilon}$ in the fully reduced
phase space action (\ref{eq:fully-reduced-action}) by~$P^{-\epsilon}$,
defined by the contour integral, gives the action the imaginary
contribution 
$\pm i \epsilon \pi \int {\kappa}^{-1} dM 
= 
\pm i \epsilon \pi \int d(\phi_H/G) 
= 
\pm i (\epsilon \hbar/2)  \int d S_{\mathrm{BH}} 
$, 
where we have used the identity $dM =
\kappa d(\phi_H / G)$ and the consequence that the Bekenstein-Hawking
entropy is given by $S_{\mathrm{BH}} = 2\pi \phi_H / (\hbar G)$. This
calculation has some similarity to the tunneling analyses that have
led to the Bekenstein-Hawking entropy and to corrections thereof in
the contexts of
\cite{observables,geg-kunst-tunneling,parikh-wilczek,paddy-etal,%
vagenas-etal,nadalini-etal,kerner-mann,akh2-sing}, 
including the numerical factor 
$\frac12$, which leads to the expected exponential probability factor 
$\exp \bigl(\mp \epsilon \int d S_{\mathrm{BH}} \bigr)$. For our
system, this probability factor however equals unity when evaluated on
a classical solution, since $M$ and $S_{\mathrm{BH}}$ then do not evolve in
time. We are therefore not aware of ways to develop this observation
further in the present context.

Consider then the quantum theory of Section 
\ref{sec:quantization} with given~$\epsilon$.
The operator counterpart of $f P^{-\epsilon}$ satisfies 
\begin{align}
\widehat{fP^{-\epsilon}} \, \Phi_M^{\epsilon}
&= 
\widehat{fP^{\epsilon}} \, \Phi_M^{\epsilon}
\nonumber
\\
&\phantom{=}
+ 2 \epsilon f(M) 
\Phi_M^{\epsilon}
\int
dx \, 
\frac{\sqrt{2lG M j}
+ 
\left(
\sqrt{j^2 - \hbar^2G^2/4}  - j
\right)
}
{j- 2lG M} , 
\label{eq:fPquantum-eps}
\end{align}
where we have explicitly included the relevant superscripts 
$\pm\epsilon$ on the
states and the operators. 
Compared with the classical relation~(\ref{eq:fPdiff}), 
$\widehat{fP^{-\epsilon}}$ thus contains an
additional term, which can be interpreted as a quantum correction. 
Taking  
the integral in (\ref{eq:fPquantum-eps})
to be defined as a contour integral and assuming $f$ to be
real-valued, we find 
\be
\left[ \mathrm{Im} \,  
\bigl( \widehat{fP^{-\epsilon}} 
\bigr) 
\right] 
\Phi_M^{\epsilon}
= 
\mp 
\epsilon \pi f(M)
\, 
\widehat{\kappa^{-1}}
\, 
\Phi_M^{\epsilon} ,  
\label{eq:fPquantum-eps-diff-imag}
\ee
where the operator $\widehat{\kappa^{-1}}$ is defined by 
\be 
\widehat{\kappa^{-1}}
\, 
\Phi_M^{\epsilon}
: = 
\frac{1}{\kappa(M)} 
\, 
\Theta \! 
\left( 
1 - \frac{\hbar^2}{16 l^2 M^2}
\right)
\sqrt{1 - \frac{\hbar^2}{16 l^2 M^2}} \; 
\Phi_M^{\epsilon} , 
\label{eq:invkappa-qcorr}
\ee
$\Theta$ being the Heaviside function. 
Comparison of (\ref{eq:fPminus-imagpart-solution})
and (\ref{eq:fPquantum-eps-diff-imag}) 
shows that we may regard 
$\widehat{\kappa^{-1}}$ 
as the inverse surface gravity operator in the quantum theory. 

That $\widehat{\kappa^{-1}}$ differs from multiplication by the
classical inverse surface gravity is a consequence
of the fluctuations off the classical constraint surface that are
present in our Dirac quantization of the Hamiltonian constraint.
$\widehat{\kappa^{-1}}$ is close to the classical inverse surface
gravity for $M \gg
\hbar/(4l)$, but the difference becomes significant at the Planck
scale, and $\widehat{\kappa^{-1}}$ vanishes on all states whose
support is at $M \le \hbar/(4l)$.

\section{Conclusions}
\label{sec:conclusions}

In this paper we have presented a Dirac quantization of generic
single-horizon black holes in two-dimensional dilaton gravity, working
under boundary conditions that allow the spatial surfaces to extend
from a singularity to an infinity and eliminating the spatial
reparametrization freedom by a spatial gauge choice at the classical
level. The Hamiltonian constraint that remains was quantized in a
metric representation. After finding a vector space of ADM mass
eigenstate solutions to the quantum constraint, we transformed to a
representation that allowed the mass spectrum to become continuous,
and we chose the inner product by requiring self-adjointness of a time
operator that is affinely conjugate to the ADM mass. 

As the classical theory does not have local propagating degrees of
freedom, one might not expect the quantum theory to have observables
that correspond to localised geometric quantities in the
spacetime. However, both the classical theory and the quantum theory
were constructed under boundary conditions that distinguish future and
past horizons, and we used this distinction to identify in the quantum
theory an operator that corresponds to the inverse surface gravity of
the horizon. The difference from the classical surface gravity is small
for large ADM masses but becomes significant when the ADM mass
approaches the Planck mass, and below (a numerical multiple of) the
Planck mass the inverse surface gravity operator is identically
vanishing. 

For technical concreteness, we focused on boundary conditions under
which the spatial surfaces asymptote to the PG foliation both at
the singularity and at the infinity. While the technicalities of the
spatial falloff depend on this choice, both the classical and the
quantum analysis has a conceptually straightforward generalization to
any asymptotics that retains the notion of freely specifiable
asymptotic Killing time evolution. The only significant change in the
classical observables is that $\Pi_{\prmassfunction}$
(\ref{eq:Pi-prmassf-def}) contains an additional term, which accounts
for the transition from the PG time coordinate in
(\ref{eq:Tprimesol1}) to the time coordinate that determines the new
asymptotics. This term depends on $j$ and any
functions of $\phi$ that are introduced to specify the new foliation,
but it depends on $\rho$ and $\Pi_\rho$ only through the
combination~$\prmassfunction$. Assuming that we work with smooth
foliations, the new term is also smooth. The operator
$\widehat{\Pi_{\prmassfunction}}$ (\ref{eq:Pi-prmass-redop}) contains
then the same additional term, but since this term is smooth, there is
no change in the factor ordering parameter~(\ref{eq:eta-raw}), and
consequently there is no change in the singular part
in~(\ref{eq:fPquantum-eps}). Hence the inverse surface gravity
operator (\ref{eq:invkappa-qcorr}) is unchanged. Note that we can in
particular choose the foliation near infinity to be asymptotic to
the surfaces of constant Schwarzschild time, in which case the
concerns of Section \ref{sec:cross-qhor} about the convergence of the
integrals at $x\to\infty$ do not arise.

Similarly for technical concreteness, we focused on the spatial gauge
choice (\ref{eq:gauge-condition}) when eliminating the spatial
reparametrization freedom in the classical theory. There is a
straightforward generalization to gauge conditions of the form
\be
l\phi' - g(\phi) =0 , 
\label{eq:gauge-condition-gen}
\ee
provided the positive gauge fixing function $g$ 
allows the 
spatial hypersurfaces to extend from a singularity to an infinity and
suitable falloff conditions to be imposed. Assuming this is the case,
the significant changes are that in the classical theory 
(\ref{eq:Pi-prmassf-def}) 
is replaced by 
\be
\Pi_{\prmassfunction} := \frac{lG \Pi_\rho
-\epsilon g \sqrt{2lG \prmassfunction /j}}
{j- 2lG \prmassfunction} , 
\label{eq:Pi-prmassf-gendef}
\ee
and in the quantum theory (\ref{eq:nusquared-def}) is replaced by 
\be
\nu^2 = \frac{1}{4} - \frac{g^2}{\hbar^2 G^2} . 
\label{eq:nusquared-gendef}
\ee
The inverse surface gravity operator then reads
\be 
\widehat{\kappa^{-1}}
= 
\frac{1}{\kappa(M)} 
\, 
\Theta \! 
\left( 
1 - \frac{\hbar^2 G^2}{4 {\bigl[g(\phi_M)\bigr]}^2 }
\right) 
\sqrt{1 - \frac{\hbar^2 G^2}{4 {\bigl[g(\phi_M)\bigr]}^2 }} , 
\label{eq:invkappa-qcorr-gengauge}
\ee
where $\phi_M$ is the solution to 
\be
j(\phi_M) = 2 l G M . 
\label{eq:phiM-def}
\ee
The inverse surface
gravity operator therefore depends on the choice of~$g$. To discuss
this dependence further, one would need to develop a
more quantitative control of the class of $g$s that are compatible
with the boundary conditions of the classical theory.

Three points should be emphasized. First, the difference between the
inverse surface gravity operator (\ref{eq:invkappa-qcorr-gengauge})
and the classical inverse surface gravity $\kappa^{-1}(M)$ arises
because the Hamiltonian constraint was \emph{not\/} eliminated at the
classical level but instead quantized in the Dirac sense as an
operator. The regularity of the quantum observables across the future
and past horizons was formulated in a way that hinges on the
fluctuations off the classical constraint surface, and it was the
distinction between regularity across the future horizon and past
horizon that led to the identification of the inverse surface gravity
operator.

Second, we chose to quantize the partially reduced theory in a
`metric' representation. We introduced on the classical phase space a
chart in which the variables are closely related to the local
spacetime geometry, and the geometry of this chart then inspired the
technical input in our quantization, leading in particular to the
notion of regularity of the quantum observables across the Killing
horizon. In comparison, it is possible to introduce in the (fully)
unreduced classical theory a phase space chart that separates the
constrained and unconstrained degrees of freedom: the unconstrained
coordinates can be chosen as the ADM mass and the Killing time
difference between the asymptotic ends of the spatial surfaces,
whereas all the remaining information about the embedding of the
spatial surfaces in the spacetime becomes encoded in the pure gauge
degrees of freedom
\cite{kas-thie1,kas-thie2,kuchar,varadarajan}. 
Quantum theories whose technical input is inspired by such a chart
have been 
given \cite{kas-thie1,kas-thie2,kuchar,varadarajan}, and
these quantum theories can be specialized to boundary conditions that
place one end of the spatial surfaces at a Killing horizon
\cite{louko-whiting,bose-etal,louko-winters,louko-simon-winters,%
kunstatter-petryk-shelemy,kiefer-louko,%
medved-kunstatter,medved-revisit}. However, the geometry of such a
phase space chart does not appear to suggest a horizon-crossing
regularity condition in the quantum theory, and introducing an
operator related to surface gravity would require other input. While
it is well known that inequivalent quantum theories can arise from
quantizations that draw their input from different phase space charts,
the specific issue here may be related to the observation that
geometrically nontransparent quantum variables can produce large
quantum fluctuations in the spacetime geometry
\cite{ashtekar-fluctuations,gambini-pullin,vara-CGHS}.

Third, the inverse surface gravity operator $\widehat{\kappa^{-1}}$
(\ref{eq:invkappa-qcorr-gengauge}) depends on the partial gauge-fixing
condition (\ref{eq:gauge-condition-gen}) in a way that has a geometric
meaning. By~(\ref{eq:phiM-def}), $\phi_M$ is the value of $\phi$ on
the horizon of the classical solution with
mass~$M$. $\widehat{\kappa^{-1}}$~hence knows how the gauge choice
makes the spatial surfaces cross the horizon but does not know what
the surfaces do elsewhere. On the one hand, this is pleasing: the
formalism relates the quantum-corrected surface gravity to the
embedding of the spatial surfaces precisely where the surfaces cross
the horizon. On the other hand, what is unsatisfactory is that the
gauge choice was made already at the classical level. One would like
first to quantize the theory in a gauge-invariant way, and if
operators that pertain to specific foliations are desired, to
introduce such operators only in the already-quantized
theory. Unfortunately, our quantization technique relied in an
essential way on the decoupling of the spatial points in the mass
operator~(\ref{eq:prmassoperator-def}), and this decoupling only arose
because the spatial diffeomorphism constraint was eliminated
classically. If one attempted to treat also the spatial diffeomorphism
constraint as a quantum constraint, one new issue would be how to
preserve the constraint algebra in the quantum
theory~\cite{kuchar-winnipeg,lew-mar,GLMP}.

Given a function on the phase space
of the fully reduced classical theory, one can
explore the options of quantizing this function in our 
quantum theory via some interpretation of the rule ``$P
\mapsto - i \hbar\partial_M$ modulo factor ordering''. 
However, there is no guarantee that a reasonable interpretation can be
found for all functions of geometric interest. As an example, fix
$\epsilon$ and consider the function 
\be
\lambda(M,P) := \epsilon e^{-\epsilon \kappa(M) P} . 
\label{eq:lambda-MP}
\ee
By solving the geodesic equation on the horizon, it can be verified
that $\lambda(M,P)$ is an affine parameter for the null geodesic that
straddles the horizon, in a foliation that coincides with the PG
coordinates except near the singularity and is at the singularity
asymptotic to a single surface of constant PG time. Note that this
means $\sigma_0=0$ and $\sigma_\infty = 1$ in
(\ref{eq:smallphi-falloff}) and~(\ref{eq:largephi-falloff}). The
affine parameter increases to the future and has been normalised so
that it vanishes at the bifurcation point and equals $\epsilon$ on the
surface $P=0$. Now, if one had a self-adjoint operator version
of~$\kappa(M)P$, the operator exponential in (\ref{eq:lambda-MP})
could be defined by spectral analysis. Suppose for concreteness that
$\kappa(M)$ is proportional to $M^{1+2\gamma}$ with
$\gamma\in\mathbb{R}$, which covers in particular symmetry-reduced
gravity. The substitution $M^{1+2\gamma} P
\mapsto - i \hbar M^{1+\gamma} \partial_M M^{\gamma}$ yields a
symmetric operator, but analysis of the deficiency indices
\cite{thirring} shows that this operator has 
no self-adjoint extensions except when $\gamma=0$, and $\gamma=0$ is
not consistent with the assumed asymptotic structure of the spacetime
at $\phi\to\infty$. We have therefore not found a reasonable
quantization of the affine parameter of the horizon in the present
formalism.

As the classical system has no local propagating degrees of freedom,
it seems unlikely that our inverse surface gravity operator could be
used to make predictions in terms of Hawking radiation or black hole
entropy. We expect however a number of the features of our quantum
theory to be generalizable upon inclusion of matter with local
dynamics, in particular the way how regularity of quantum operators
across the horizon is defined in the presence of quantum fluctuations
off the classical constraint surface. Given a suitable adaptation of
our boundary conditions to accommodate local dynamics~\cite{dgk}, 
it may thus be possible to generalize our techniques to study both
Hawking radiation and singularity formation in the quantum theory.

\section*{Acknowledgements}

We thank 
Ramin Daghigh, 
Jack Gegenberg, 
Viqar Husain
and 
Oliver Winkler 
for 
useful discussions
and an anonymous referee for helpful comments. 
GK~thanks the University of Nottingham 
and JL thanks the Perimeter Institute for 
Theoretical Physics for hospitality. 
GK~was supported in part by the 
Natural Sciences and Engineering 
Research Council of Canada. 
JL~was supported in part by PPARC grant PP/D507358/1.


\end{document}